\documentclass{elsart5p}

\usepackage{graphics}
\usepackage{graphicx}
\usepackage{amssymb}
\begin{document}

\begin{frontmatter}



\title{Quasi-quartet crystal electric field ground state\\ in a tetragonal CeAg$_2$Ge$_2$ single crystal}

%\author[AA]{A. Name1\corauthref{Name1}},
%\ead{name1@uni1.ac.xy}
%\author[AA]{B. Name2},
%\author[AA,BB,CC]{C. Name3}

%\address[AA]{Address 1}
%\address[BB]{Address 2}
%\address[CC]{Address 3}

\author{A. Thamizhavel\corauthref{Thamizhavel}},
\ead{thamizh@tifr.res.in}
\author{R. Kulkarni},
\author{S. K. Dhar}
\corauth[Thamizhavel]{Corresponding author. Tel: (81)22-2280-4556}
\address{Department of condensed matter physics and materials science, \\Tata institute of fundamental research, Colaba, Mumbai 400 005, India}

\begin{abstract}
% Text of abstract
We have successfully grown the single crystals of CeAg$_2$Ge$_2$, for the first time, by flux method and studied the anisotropic physical properties by measuring the electrical resistivity, magnetic susceptibility and specific heat.  We found that CeAg$_2$Ge$_2$ undergoes an antiferromagnetic transition at $T_{\rm N}$ = 4.6~K.  The electrical resistivity and susceptibility data reveal strong anisotropic magnetic properties.  The magnetization measured at $T$~=~2~K exhibited two metamagnetic transitions at $H_{\rm m1}$ = 31~kOe and $H_{\rm m2}$ = 44.7~kOe, for $H~\parallel$~[100] with a saturation magnetization of 1.6~$\mu_{\rm B}$/Ce.  The crystalline electric field (CEF) analysis of the inverse susceptibility data reveals that the ground state and the first excited states of CeAg$_2$Ge$_2$ are closely spaced indicating a quasi-quartet ground state.  The specific heat data lend further support to the presence of closely spaced energy levels. 
\end{abstract}


\end{frontmatter}

% main text

%\section{Introduction}
%\label{}
Compounds crystallizing in the ThCr$_2$Si$_2$ type structure are the most extensively studied among the strongly correlated electron systems.  A wide range of compounds crystallize in this type of tetragonal crystal structure and exhibit novel physical properties.  Some of the prominent examples include the first heavy fermion superconductor CeCu$_2$Si$_2$, pressure induced superconductors like CePd$_2$Si$_2$, CeRh$_2$Si$_2$, CeCu$_2$Ge$_2$, unconventional metamagnetic transition in CeRu$_2$Si$_2$ etc.  CeAg$_2$Ge$_2$ also crystallizes in the tetragonal  ThCr$_2$Si$_2$ type crystal structure.  Previous reports of CeAg$_2$Ge$_2$ were on polycrystalline samples and there were conflicting reports on the antiferromagnetic ordering temperature~\cite{Rauchschwalbe, knopp, cordruwisch}.  Furthermore, the ground state properties of CeAg$_2$Ge$_2$ are also quite intriguing.  Neutron scattering experiments on a polycrystalline sample could detect only one excited state at 11~meV indicating that the ground state and the first excited states are closely spaced. In order to study the anisotropic physical properties and to study the crystalline electric field ground state, we have grown the single crystals of CeAg$_2$Ge$_2$.   

\begin{figure}[!]
\begin{center}
\includegraphics[width=0.45\textwidth]{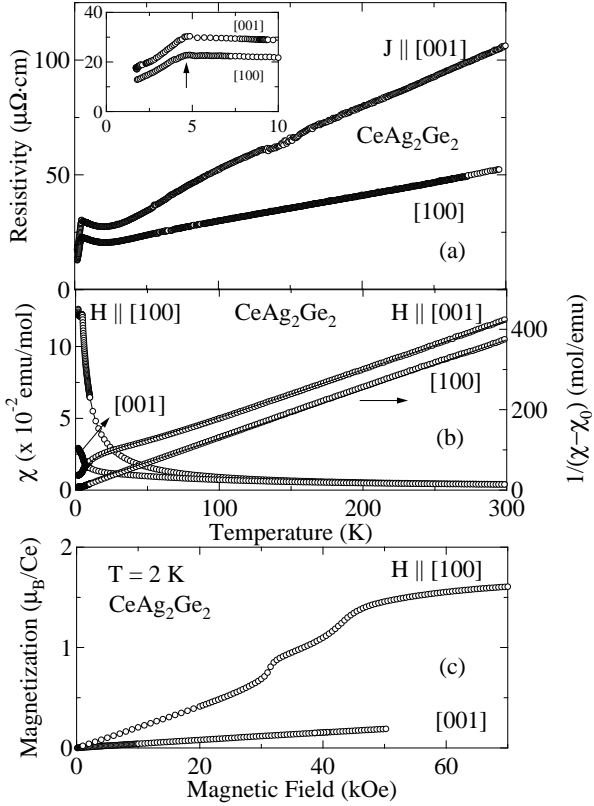}
\end{center}
\caption{(a) The temperature dependence of electrical resistivity of CeAg$_2$Ge$_2$, inset shows the low temperature part, (b) Temperature dependence of the magnetic susceptibility together with inverse magnetic susceptibility plot, solid lines indicate the CEF fitting and (c) Magnetization of CeAg$_2$Ge$_2$ measured at $T$~=~2~K.} \label{Fig1}
\end{figure}

%\section{Experiments and Results}
Single crystals of CeAg$_2$Ge$_2$ were grown by self flux method, using Ag:Ge (75.5: 24.5) binary alloy, which forms an eutectic at 650~$^\circ$C, as flux. The details about the crystal growth process are given elsewhere~\cite{thamizh}.  Figure~\ref{Fig1}(a) shows the temperature dependence of electrical resistivity of CeAg$_2$Ge$_2$ for the current direction parallel to both [100] and [001] directions.  There is a large anisotropy in the electrical resistivity.  The electrical resistivity shows a shallow minimum at 20~K, marginally increases  with decrease in temperature down to 4.6~K.   With further decrease in the temperature the electrical resistivity drops due to the reduction in the spin-disorder scattering caused by the antiferromagnetic ordering of the magnetic moments, as seen in the inset of Fig.~\ref{Fig1}(a).  The antiferromagnetic transition can be clearly seen at 4.6~K as indicated by the arrow in the figure.  

Figure~\ref{Fig1}(b) shows the temperature dependence of the magnetic susceptibility along the two principle directions.  As can be seen from the figure there is a large anisotropy in the susceptibility due to tetragonal crystal structure.  The high temperature susceptibility does not obey the simple Curie-Weiss law; on the other hand it can be very well fitted to a modified Curie-Weiss law which is given by $\chi = \chi_{\rm 0} + \frac{C}{T-\theta_{\rm p}}$, where $\chi_{\rm 0}$ is the temperature independent part of the magnetic susceptibility and $C$ is the Curie constant.  The value of $\chi_{\rm 0}$ was estimated to be 1.33~$\times$~10$^{-3}$ and 1.41~$\times$~10$^{-3}$~emu/mol for $H~\parallel~$[001] and [100], respectively such that an effective moment of 2.54~$\mu_{\rm B}$/Ce is obtained for temperatures above 100~K.  In order to perform the CEF analysis of the susceptibility data, we plotted the inverse susceptibility as 1/($\chi - \chi_{\rm 0}$) versus $T$. The solid line in figure~\ref{Fig1}(b) are the fitting to the inverse susceptibility with the CEF Hamiltonian given by $\mathcal{H}_{\rm CEF} = B_{2}^{0}O_{2}^{0} + B_{4}^{0}O_{4}^{0} + B_{4}^{4}O_{4}^{4}$, where $B^m_l$ and $O^m_l$ are CEF parameters and the Stevens operators respectively.  The level splitting energies are estimated to be $\Delta_{\rm 1}$~=~5~K and $\Delta_{\rm 2}$~=~130~K.  It may be noted that the first excited state is very close to the ground state indicating that the ground state is a quasi-quartet state.  Figure~\ref{Fig1}(c) shows the field dependence of magnetization at 2~K.  For $H~\parallel$~[100], the magnetization increases linearly with the field and exhibit two metamagnetic transition at $H_{\rm m1}$ = 31~kOe and $H_{\rm m2}$ = 44.7~kOe before it saturates at 1.6~$\mu_{\rm B}$/Ce at 70~kOe, indicating the easy axis of magnetization.  On the other hand the magnetization along [001] is very small and varies linearly with field reaching a value of 0.32~$\mu_{\rm B}$/Ce at 50~kOe.

\begin{figure}
\begin{center}
\includegraphics[width=0.45\textwidth]{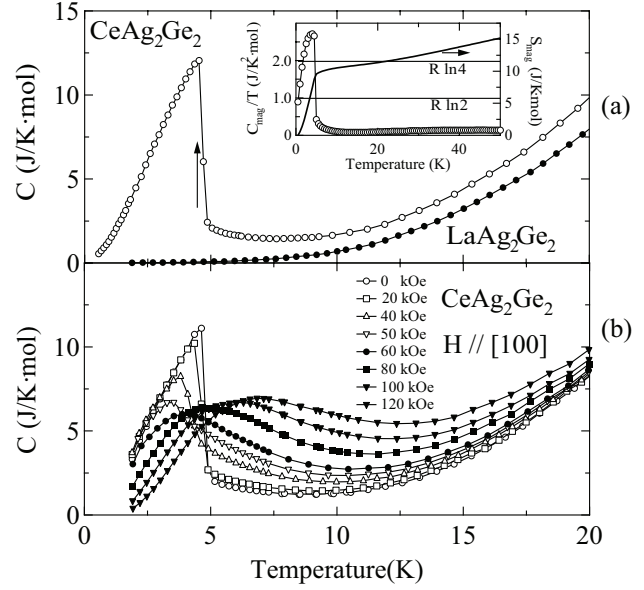}
\end{center}
\caption{(a)  Temperature dependence of the specific heat of CeAg$_2$Ge$_2$ and LaAg$_2$Ge$_2$.  The inset shows the magnetic entropy.  (b)  The field dependence of the specific heat of CeAg$_2$Ge$_2$ for the field applied along the easy axis of magnetization, namely [100].} \label{Fig2}
\end{figure}

Figure~\ref{Fig2}(a) shows the temperature dependence of the specific heat of CeAg$_2$Ge$_2$ together with the specific heat of a reference sample LaAg$_2$Ge$_2$.  The antiferromagnetic ordering is manifested by the clear jump in the specific heat at $T_{\rm N}$ = 4.6~K as indicated by the arrow.  The inset of Fig.~\ref{Fig2}(a) shows the $C_{\rm mag}/T$ versus T together with the magnetic entropy.  The entropy reaches $R$~ln~4 not too far away from the magnetic ordering temperature leading to the conclusion that the ground state and the first excited states are closely spaced or nearly degenerate, thus corroborating our CEF analysis of the inverse susceptibility data.  Figure~\ref{Fig2}(b) shows the field dependence of the specific heat for the field applied parallel to the easy axis of magnetization namely [100].  With the increase in the magnetic field the N\'{e}el temperature decreases  and the antiferromagnetic ordering apparently vanishes at a critical field of 50~kOe indicating a possibility of a field induced quantum critical point in this compound.  However, further low temperature measurements are necessary to confirm this.  

In summary, we have successfully grown the single crystals of CeAg$_2$Ge$_2$ by the flux method.  CeAg$_2$Ge$_2$ orders antiferromagnetically at $T_{\rm N}$ = 4.6~K.  The CEF analysis of the inverse susceptibility data indicate the ground state and the first excited states are closely spaced.  The heat capacity data support this quasi-quartet ground state.  Furthermore, the heat capacity in applied magnetic fields revealed that the N\'{e}el temperature vanishes at a critical field of 50~kOe indicating a possible field induced quantum critical point in this compound.

%\begin{table}
%\caption{Lattice parameters of ABC}
%\begin{center}
%\begin{tabular}{|c|c|c|c|c|}
%\hline 1 & 2 & 3 & 4 & 5 \\
%\hline 1.23 & 2.34 & 3.45 & 4.56 & 5.67 \\
%\hline
%\end{tabular}
%\end{center}
%\end{table}

%\section{Summary}

%\section{Acknowledgement}


\begin{thebibliography}{99}

%\bibitem{Paper1} A. Author1, Adv. Phys. {\bf 100} (2006) %111.

\bibitem{Rauchschwalbe}R. Rauchschwalbe et al.,  J. Less Common. Metals {\bf 111}, (1985) 265.

\bibitem{knopp}G. Knopp et al.,  J. Magn. Magn. Mater. {\bf 63~\&~64}, (1987) 88.

\bibitem{cordruwisch}E. Cordruwish et al.,  J. Phase Equilibria {\bf 20},  (1999) 407.

\bibitem{thamizh}A. Thamizhavel et al., Phys. Rev. B (2007) to be published

%\bibitem{Paper2} B. Author1 et al., Phys. Rev. B {\bf 75} %(2007) 222.

\end{thebibliography}
\end{document}